\documentclass[a4paper,11pt,reqno]{amsart}
\usepackage{amssymb}
\usepackage{graphicx}
\numberwithin{equation}{section}
\swapnumbers
\theoremstyle{plain}
\newtheorem{thm}{Theorem}[section]
\newtheorem{lem}[thm]{Lemma}
\newtheorem{prop}[thm]{Proposition}
\newtheorem{cor}[thm]{Corollary}
\theoremstyle{definition}
\newtheorem*{Def*}{Definition}
\newtheorem*{rems*}{Remarks}
\newtheorem*{rem*}{Remark}
\providecommand{\C}[1]{\mathcal{#1}}
\providecommand{\D}[1]{\mathbb{#1}}

\providecommand{\F}[1]{\mathfrak{#1}}

\newcommand{\NN}{{\mathbb N}}    
\newcommand{\RR}{{\mathbb R}}    
\newcommand{\CalF}{\mathcal{F}}
\newcommand{\dell}{\mathbb{D}}
\newcommand{\TT}{\mathbb T}

\DeclareMathOperator{\supp}{supp}
\begin{document}
%-------------------------------------------------------------%
\title[Generically singular continuous spectrum]
       {Generic sets in spaces of measures and generic
singular continuous spectrum for Delone Hamiltonians}
\author[D.~Lenz]{Daniel Lenz$^\dagger$}
\author[P.~Stollmann]{Peter Stollmann$^\dagger$}
\address{$^{\dagger}$  Fakult\"at f\"ur Mathematik,
          Technische Universit\"at, 09107 Chemnitz, Germany}

%-------------------------------------------------------------%
\date{01.04.2004}
%-------------------------------------------------------------%

\begin{abstract}  
We show that geometric disorder leads to 
purely singular continuous spectrum generically. 

The main input is a result of Simon known as the
``Wonderland theorem''. Here, we provide an alternative approach and
actually a slight strengthening by showing that various sets of
measures defined by  regularity properties are generic in the set of
all measures on a locally compact metric space.

As a byproduct we obtain that a generic measure on euclidean space
is singular continuous.

\end{abstract}
%-------------------------------------------------------------%
\maketitle
%-------------------------------------------------------------%
\section{Introduction}                             \label{s1}
In this paper we study the spectral type of continuum Hamiltonians
describing solids with a specific form of disorder that could be
called geometric disorder. The corresponding Hamiltonians are defined
using Delone sets (uniformly discrete and uniformly dense subsets
of euclidean space) in the following way: the ions of a solid are assumed
to be distributed in space according to points of a Delone set. We fix
an effective potential $v$ for each of the ions and consider the
effective Hamiltonian
$$ H(\omega) := - \Delta + \sum_{x\in \omega} v( \cdot -x)$$ for every
Delone set $\omega \in \dell_{r,R}$ (see below for precise
definition). We can now show that under some mild assumptions
concerning $v,r,R$ there exists a dense $G_\delta$-set of $\omega$'s
for which $H(\omega)$ exhibits a purely singular continuous spectral
component.

\smallskip

Let us put this result into perspective. Two extremal forms of
geometrically disordered sets are periodic sets and highly random sets
with points distributed according to, say, a Poisson law. In the
periodic case, it is known that the spectrum is purely absolutely
continuous. For the highly random case some pure point spectrum is
expected. The region in between these two extremes contains the case of
aperiodic order. This form of disorder has attracted much attention
recently due to the discovery of solids exhibiting this kind
of order (see e.g. \cite {Moo,BM} for reviews and further references). 
These solids are called quasicrystals. For them singularly continuous 
spectrum is assumed to occur.

This picture is well substantiated by one-dimensional
investigations \cite{BS}. However, in the higher dimensional case essentially
nothing is known (except of course absolute continuity of the spectrum for
the periodic case).

Thus, our result provides a first modest step in the investigation of
geometric disorder in the higher dimensional case.

\smallskip

The main tool we use goes back to Simon's Wonderland theorem. It
basically tells us that certain sets of operators defined by spectral
types are in fact regular in the sense that they are
$G_\delta$-sets. If, by chance, one can prove denseness of these
sets, then their intersection is dense as well. For instance, in the
situation indicated above, we can prove that a dense set of $\omega$'s
leads to continuous spectral measures (actually, even absolutely
continuous spectral measures) and a dense set of $\omega$'s leads to
singular spectral measures (in some energy interval). Since both these
sets turn out to be $G_\delta$, we get a dense $G_\delta$-set of
$\omega$'s such that the spectrum of $H(\omega)$ is singular
continuous.

We need a slight generalization of Simon's results. More importantly,
we provide a new proof of his result which gives some new
insights. Namely, we consider the relevant sets directly in spaces of
measures and show that the singular (w.r.t. some comparison measure)
measures form a $G_\delta$-set in the space of Borel measures on a
nice metric space. The same holds for the continuous measures. By
standard continuity arguments, this regularity can be pulled back to
spaces of operators.

Since we work directly in spaces of measures we obtain as a byproduct
of our investigation that the set of singular continuous measures on
euclidean space is a dense $G_\delta$-set in the space of
measures. Although this could certainly be expected in view of all
the strange things that typically happen as a consequence of Baire's
theorem we are not aware of a proof of this fact.

\bigskip
\medskip

\noindent\textbf{Acknowledgment:}\\
\noindent Our collaboration has been supported in part by the DFG
in the priority program ``Quasicrystals''. Part of the results presented 
here have been announced in \cite{LS}. Useful comments of J. Voigt are
gratefully acknowledged. They led to more elegant proofs in Section 2.

%-------------------------------------------------------------%
\section{Back to Wonderland}       \label{s2}
The following result is the main abstract input to our proof
of generic appearance of a purely singular continuous component
in the spectrum of certain Delone Hamiltonians.

It is a soft result that basically follows from Baire's Theorem
and only a minor strengthening of Simon's ``Wonderland Theorem''
from \cite{si}. In order to formulate it efficiently, let us 
introduce the following notation:

For a fixed separable Hilbert space $\F{H}$ consider the
space $\F{S}=\F{S}(\F{H})$ of self-adjoint operators in $\F{H}$. 
We endow $\F{S}$ with the \emph{strong resolvent topology} $\tau_{srs}$,
the weakest topology for which all the mappings
$$
\F{S}\to \D{C}, A\mapsto (A+i)^{-1}\xi \quad (\xi\in\F{H})
$$
are continuous. Therefore, a sequence $(A_n)$ converges to
$A$ w.r.t. $\tau_{srs}$ if and only if 
$$
(A_n+i)^{-1}\xi\to (A+i)^{-1}\xi
$$
for all $\xi\in\F{H}$.

We denote by $\sigma_*(A)$ for $*=c,s,ac,sc,pp$ the continuous,
singular, absolutely continuous, singular continuous and pure
point spectrum, respectively.

\begin{thm}\label{thm2.1}
Let $(X,\rho)$ be a complete metric space and $H: (X,\rho)\to 
(\F{S},\tau_{srs})$ 
a continuous mapping. Assume that, for an open set $U\subset\D{R}$,
\begin{itemize}
\item[\textrm{(1)}] the set $\{ x\in X |\; \space  \sigma_{pp}(H(x))\cap U
=\emptyset\}$ is dense in $X$,
 
\item[\textrm{(2)}] the set $\{ x\in X |\; \space  \sigma_{ac}(H(x))\cap U
=\emptyset\}$ is dense in $X$,

\item[\textrm{(3)}] the set $\{ x\in X |\; \space   U\subset\sigma(H(x))\}$ 
is dense in $X$.
\end{itemize}
Then, the set
$$
\{ x\in X |\; \space   U\subset\sigma(H(x)), \sigma_{ac}(H(x))\cap U
=\emptyset, \sigma_{pp}(H(x))\cap U
=\emptyset\}
$$
is a dense $G_\delta$-set in $X$.
\end{thm}
  
\begin{rem*}
If the mapping $H$ in the above theorem happens to be injective,
then its range $H(X)$ can be endowed with the metric $\rho$ given on
$X$. In this case, $H(X)$ is a regular metric space of operators
in the sense of \cite{si} and Theorem 2.1 of the latter paper gives
the result.
\end{rem*}
As can be seen from this remark, the above theorem is only slightly
stronger than Simon's result.  Our proof, however, is somewhat
different. It is based on the fact that certain sets in spaces of
measures are $G_\delta$'s. In that sense we get some extra information
in comparison with \cite{si}. The basic idea is that once one has
established the $G_\delta$-property for the set of all measures that
are purely continuous and purely singular respectively, this
regularity can be pulled back to obtain the $G_\delta$-property for
the sets appearing in assumptions (1) and (2) of the theorem. Clearly
this gives the asserted denseness of the intersection of these
sets. Since the set from (3) is $G_\delta$ as well, the set appearing
in the assertion is the intersection of three dense $G_\delta$-sets
and dense as such.

We now start our program of proof by studying certain subsets of the
set of positive, regular Borel measures $\C{M}_+(S)$ on some locally
compact, $\sigma$-compact, separable metric space $S$. Of course,
$\C{M}_+(S)$ is endowed with the weak topology from $C_c(S)$, also
called the vague topology.  We refer the reader to \cite{bau} for
standard results concerning the space of measures. In particular, we
note that the vague topology is metrizable  such that $\C{M}_+(S)$
becomes a complete metric space, as we consider second countable
spaces.  For the application we have in mind, $S$ is just the open
subset $U$ of the real line that appears in the theorem.

We call a measure $\mu\in\C{M}_+(S)$ \emph{diffusive} or \emph{continuous}
if its \emph{atomic} or \emph{pure point part} vanishes, i.e. if
$\mu(\{ x\})=0$ for every $x\in S$. (We prefer the former terminology
in the abstract framework and the latter for measures on the real line.)
Two measures are said to be \emph{mutually singular}, $\mu\perp\nu$,
if there exists a set $C\subset S$ such that $\mu(C)=0=\nu(S\setminus C)$.
We have
\begin{thm}\label{thm2.2}
Let $S$ be as above.
\begin{itemize}
\item[\textrm{(1)}] The set $\{ \mu\in \C{M}_+(S)|\;   \mu\mbox{  is
diffusive}\}$ is a $G_\delta$-set in $\C{M}_+(S)$.
 
\item[\textrm{(2)}] For any  $\lambda\in\C{M}_+(S)$,
the set $\{ \mu\in \C{M}_+(S)|\;   \mu\perp\lambda\}$ is a $G_\delta$-set in $\C{M}_+(S)$.
 
\item[\textrm{(3)}] For any closed $F\subset S$ the set $\{ \mu\in \C{M}_+(S)|\;  
F\subset\supp (\mu)\}$ is a $G_\delta$-set in $\C{M}_+(S)$.
\end{itemize}
\end{thm}
\begin{proof}[Proof of \emph{Theorem \ref{thm2.2} (3):}]
First consider the case that $F=\{ x\}$. Choose a basis $(V_n)_{n\in
\D{N}}$ of open neighborhoods of $x$. Then
$$
\{ \mu\in \C{M}_+(S)|\;  
x\in\supp (\mu)\}^c=\bigcup_{n\in\D{N}}\{ \mu\in \C{M}_+(S)|\;  \mu(V_n)=0\}
$$
is an $F_\sigma$, since  $\{ \mu\in \C{M}_+(S)|\;  \mu(V)=0\}$ is closed for
any open set $V\subset S$. Therefore $\{ \mu\in \C{M}_+(S)|\;  
x\in\supp (\mu)\}$ is a $G_\delta$.

To treat the general case, take a dense subset $\{ x_n|\;   n\in\D{N}\}$
in $F$
(which is possible since $S$ is separable.) Then
$$
\{ \mu\in \C{M}_+(S)|\;  
F\subset\supp (\mu)\} =\bigcap_{n\in\D{N}}\{ \mu\in \C{M}_+(S)|\;  
x_n\in\supp (\mu)\}$$
is a countable intersection of $G_\delta$'s, hence a $G_\delta$. 
\end{proof}

To prove parts (1) and (2) of Theorem \ref{thm2.2} we use the following
observations:

\begin{prop}\label{p2.3}
Let $\C{K}\subset \C{M}_+(S)$ be compact. Then
$$
\C{K}^{\bullet}:=\{\mu\in \C{M}_+(S)|\;  \exists\nu\in\C{K}: \nu\le\mu\}
$$
is closed.
\end{prop}
\begin{proof}
Let $(\mu_n)$ be a sequence in $\C{K}^{\bullet}$ that converges
to $\mu$. Choose $\nu_n\in\C{K}$ with $\nu_n\le\mu_n$ for $n\in\D{N}$.
By compactness, we find a converging subsequence $(\nu_{n_k})$ with limit
$\nu\in\C{K}$. For any $\varphi\in C_c(S), \varphi\ge 0$, we get
\begin{eqnarray*}
\langle \mu,\varphi\rangle &=& \lim_{k\to\infty}\langle 
\mu_{n_k},\varphi\rangle \\
&\ge& \lim_{k\to\infty}\langle 
\nu_{n_k},\varphi\rangle \\
&=& \langle \nu,\varphi\rangle
\end{eqnarray*}
so that $\mu\in\C{K}^\bullet$.
\end{proof}
\begin{prop}\label{p2.4}
Let $K\subset S$ be compact and $a>0$. Then
$$
\{  a\cdot\delta_{x}|\;   x\in K\}
$$
is compact in $\C{M}_+(S)$.
\end{prop}
The \textit{proof} is evident from the fact that $S\to\C{M}_+(S), x\mapsto\delta_x$
is continuous.
\begin{prop}\label{p2.5}
Let $\lambda\in\C{M}_+(S)$, $K\subset S$  compact and $\gamma>0$ be given.
Then
$$
\C{K}:=\{ f\cdot\lambda|\;   f\in L^2(\lambda), \| f\|_{  L^2(\lambda)} \le 1, 0\le f, \supp(f)
\subset K, \int fd\lambda\ge \gamma\}
$$
is compact.
\end{prop}
\begin{proof}
The densities considered in $\C{K}$ form a closed subset 
 of the unit ball of $L^2(K,\lambda)$. Since the latter is weakly
compact and
the mapping $L^2(K,\lambda)_+\to\C{M}_+(S), f\mapsto f\lambda$   
is w-w$^*$ -continuous we get the desired compactness.
\end{proof}
\begin{proof}[Proof of \emph{Theorem \ref{thm2.2} (1) and (2):}]

For the proof of (1) consider
$$
\C{M}_1:= \{ \mu\in \C{M}_+(S)|\;   \mu\mbox{  is
diffusive}\}^c .
$$
We want to show that $\C{M}_1$ is an $F_\sigma$. By assumption on
$S$, we find a sequence of compacts $K_n\nearrow S$ and get that
$$
\C{K}_{1,n}= 
\{  \frac{1}{n}\cdot\delta_{x}|\;   x\in K_n\}
$$
is compact by Proposition \ref{p2.4}. Proposition \ref{p2.3}
yields that $\C{F}_{1,n}=\C{K}_{1,n}^{\bullet}$ is closed.
Since
$$\C{M}_1= \bigcup_{n\in\D{N}}\C{F}_{1,n}$$
we arrive at the desired conclusion.

We show (2) with a similar argument. For $K_n$ defined as in the proof of (1), let
$$ 
\C{K}_{2,n}=\{ f\lambda|\;   f\in L^2(\lambda), \|   f\|  \le 1, 0\le f, \supp(f)
\subset K_n, \int fd\lambda\ge \frac{1}{n}\}
$$
which is compact by Proposition \ref{p2.5}. Again by Proposition
\ref{p2.3} we get that $\C{F}_{2,n}=\C{K}_{2,n}^{\bullet}$ is closed.
Since 
$$
\C{M}_2= \bigcup_{n\in\D{N}}\C{F}_{2,n}$$
is an $F_\sigma$ and 
$$
\C{M}_2^c= \{ \mu\in \C{M}_+(S)|\;   \mu\perp\lambda\} ,
$$
part (2) of Theorem \ref{thm2.2} is proven.
\end{proof}

We now pull back the regularity properties derived in Theorem 
\ref{thm2.2} to regularity properties in $\F{S}$. We denote
the spectral measure of $A\in\F{S}$ for $\xi\in\F{H}$ by
$\rho^A_{\xi}$. It is defined by
$$
\langle \rho^A_{\xi},\varphi\rangle = (\varphi(A)\xi|\;  \xi) 
\mbox{  for  }\varphi\in C_c(\D{R}) .
$$ It is easy to see that $A_n\stackrel{srs}{\to}A$ implies strong
convergence $\varphi(A_n)\to\varphi(A)$ for every $\varphi\in
C_c(\D{R})$ (see, e.g., \cite{we}, Satz 9.20) which in turn gives weak
convergence $\rho^{A_n}_{\xi}\to\rho^A_{\xi}$ for every $\xi\in\F{H}$
. Thus, for each fixed $\xi \in\F{H}$, the map $\F{S} \longrightarrow
\C{M}_+(\D{R})$, $A\mapsto \rho^A_{\xi}$, is continuous.

The spectral subspaces of $A$ are defined
by
\begin{eqnarray*}
\F{H}_{ac}(A) &=& \{ \xi\in\F{H}|\;  \rho^A_{\xi} \mbox{  is 
absolutely continuous}\}\\ 
\F{H}_{sc}(A) &=&
\{ \xi\in\F{H}|\;  \rho^A_{\xi} \mbox{  is 
singular continuous}\}, \\
\F{H}_{c}(A) &=& \{ \xi\in\F{H}|\;  \rho^A_{\xi} \mbox{  is 
 continuous}\}\\ 
\F{H}_{pp}(A)&=&\F{H}_{c}(A)^{\perp}, \F{H}_{s}(A)=\F{H}_{ac}(A)^{\perp} .
\end{eqnarray*}
These subspaces are closed and invariant under  $A$. $\F{H}_{pp}(A)$ is the
closed linear hull of the eigenvectors of $A$. Recall that the
spectra $\sigma_*(A)$ are just the spectra of $A$ restricted
to $\F{H}_{*}(A)$. Using Theorem \ref{thm2.2}, we get the following:
\begin{prop}\label{p2.6} 
Let $U\subset\D{R}$ be open and $\F{G}\subset \F{H}$ a closed
subspace. Then
\begin{itemize}
\item[\textrm{(1)}] $\{ A\in\F{S}|\;  \quad \forall \xi\in\F{G}:\quad  
\rho^A_{\xi}|_U \mbox{  is 
 continuous } \}$
is a $G_\delta$,
\item[\textrm{(2)}]  $\{ A\in\F{S}|\;  \quad \forall \xi\in\F{G}:
\quad  \rho^A_{\xi}|_U \mbox{  is 
 singular } \}$
is a $G_\delta$.
\end{itemize}
\end{prop}
\begin{proof}
First, fix $\xi\in\F{H}$. We use that the mappings $\C{M}_+(\D{R})\to
\C{M}_+(U)$, $\nu\mapsto \nu|_U$ and $\F{S}\to\C{M}_+(\D{R})$,
$A\mapsto  \rho^A_{\xi}$ are continuous. Therefore we get
that
$$
 \{ A\in\F{S}|  \rho^A_{\xi}|_U\mbox{  is continuous}\}
$$
is a $G_\delta$ by Theorem \ref{thm2.2} (1), since continuous 
is  synonymous to diffusive. In the same way
$$
 \{ A\in\F{S}|  \rho^A_{\xi}|_U\mbox{  is singular}\}
$$
is a $G_\delta$ by Theorem \ref{thm2.2} (2), since singular means 
singular with respect to the Lebesgue measure. 
Using the above
mentioned fact that the spectral subspaces are closed, for any dense 
set $\{\xi_n| n\in\D{N}\}$ we get
$$
\{ A\in\F{S}| \quad \forall \xi\in\F{G}:\quad  \rho^A_{\xi}|_U  \mbox{  is 
 continuous } \}=\bigcap_{n\in\D{N}}
\{ A\in\F{S}| \quad  \rho^A_{\xi_n}|_U\mbox{  is continuous}\} 
$$
is a $G_\delta$ as well as
$$
\{ A\in\F{S}| \quad \forall \xi\in\F{G}|
\quad  \rho^A_{\xi}| _U \mbox{  is 
 singular } \}=\bigcap_{n\in\D{N}}
\{ A\in\F{S}|\quad   \rho^A_{\xi_n}| _U\mbox{  is singular}\} .
$$
\end{proof}

For completeness sake let us reproduce the well known fact of
lower semi continuity of spectra under strong continuity; see,
e.g. \cite{si}, Lemma 1.6 or \cite{we}, Satz 9.26 (b).

\begin{prop}\label{p2.7}
Let $V\subset\D{R}$ be open. Then
$$
\{ A\in\F{S}|\;  \sigma(A)\cap V=\emptyset\}
$$
is closed.
\end{prop}
\begin{proof}
$$
\{ A\in\F{S}|\;  \sigma(A)\cap V=\emptyset\}=\bigcap_{\varphi\in C_c(V)}
\{ A\in\F{S}|\;   \varphi(A)=0\}
$$
is closed by the above mentioned continuity of $A\mapsto\varphi(A)$.
\end{proof}

We can now put all that together for the
\begin{proof}[Proof of \emph{Theorem \ref{thm2.1}:}]
By continuity of $H$ and Proposition \ref{p2.6} applied with
$\F{G}=\F{H}$ we get that the sets appearing in assumptions (1)
and (2) of the theorem are $G_\delta$'s. Choosing a countable base
$V_n, n\in\D{N}$ of $U$, we find that
$$\{ A\in \F{S}|  \quad   U\subset\sigma(A)\}^c = 
\bigcap_{n\in\D{N}} \{ A\in \F{S}|  \quad  \sigma(A)\cap V_n=\emptyset\}
$$
is an $F_\sigma$. Thus, invoking again the continuity $H$, we infer that 
the set appearing in assumption (3) of
the theorem is a $G_\delta$. 
Therefore, the asserted denseness follows by Baire's theorem, since
the set appearing in the assertion is just the intersection of the
three dense $G_\delta$'s from (1)-(3)
\end{proof}

Let us end our excursion to Wonderland by emphasizing the special role
of singular continuity witnessed above: Let $U$ be an open subset of
$\D{R}^d$.  Then, both 
$$ \C{M}_{ac} (U) := \{ \mu\in \C{M}_+(U) |  \quad 
\mu\mbox{ is absolutely continuous}\}$$ 
and 
$$ \C{M}_{pp} (U) := \{
\mu\in \C{M}_+(U) |  \quad  \mu\mbox{ is pure point}\}$$ 
are dense in
$\C{M}_+(U)$ as can be seen by standard arguments.  Thus, by Theorem
\ref{thm2.2} (1) and (2), the sets $\C{M}_{c} (U) := \{ \mu\in
\C{M}_+(U)| \quad  \mu\mbox{ is continuous}\}$ and $\C{M}_{s} (U) := \{
\mu\in \C{M}_+(U)| \quad  \mu\mbox{ is singular}\}$ are dense
$G_\delta$'s. Therefore,
  $$\C{M}_{sc} (U) :=\{ \mu\in \C{M}_+(U)| \quad  \mu\mbox{ is singular
continuous}\} = \C{M}_{c} (U)\cap \C{M}_{s} (U)$$ is a dense
$G_\delta$ by Baire's theorem. Then, another application of Baire's
theorem, shows that neither $ \C{M}_{ac} (U)$ nor $\C{M}_{pp} (U)$ can
be a $G_\delta$, as they do not intersect $\C{M}_{sc} (U)$.

Let us state the consequences for the special case of euclidean space.

\begin{cor}\label{genericsing}
Let $U$ be an open subset of $\D{R}^d$. Then the singular continuous
measures $\C{M}_{sc} (U)$ form a dense $G_\delta$ in the space
$\C{M}_+(U)$ of Borel measures.
\end{cor}
Of course, much more general spaces and reference measures can be treated:
the only important restriction is that the measures singular w.r.t the
reference measure and the diffusive measures form  dense sets.

Once more we find that the silent majority consists of rather
strange individuals. In mathematical terms, we owe this fact to
Baire and completeness. We refer to \cite{zamf1,zamf2} for a similar result
saying that most continuous monotonic functions on the real line
are not differentiable.
 Let us also mention the classical papers
\cite{ban,mazu} dealing with the lack of differentiability for a
 typical (in the sense of dense $G_\delta$'s) continuous
functions.

%%%%%%%%%%%%%%%%%
\section{Delone sets and Delone Hamiltonians}
Let us now define the operators for which we want to prove generic
singular continuous spectral components. We start by recalling what a
\emph{Delone set} is, a notion named after B.N. Delone (Delaunay),
\cite{del}. We write $U_r(x)$ and $B_r(x)$ for the open and closed
ball in $\D{R}^d$, respectively. The Euclidean norm on $\D{R}^d$ is
denoted by $\| \cdot \| $.

\begin{Def*}
A set $\omega\subset \D{R}^d$ is called an $(r,R)$-\emph{set} if
\begin{itemize}
\item $\forall x,y\in\omega, x\not= y: U_r(x)\cap U_r(y)=\emptyset$,
\item $\bigcup_{x\in\omega}B_R(x)=\D{R}^d$.
\end{itemize}
By $\dell_{r,R}(\D{R}^d)=\dell_{r,R}$ we denote the set of all
$(r,R)$-sets. We say that $\omega\subset \D{R}^d$ is a \emph{Delone
set}, if it is an $(r,R)$-set for some $0<r\le R$ so that $
\dell(\D{R}^d)=\dell=\bigcup_{0<r\le R}\dell_{r,R}(\D{R}^d)$ is the
set of all Delone sets.
\end{Def*}
Delone sets turn out to be quite useful in the description of
quasicrystals and more general aperiodic solids; see also
\cite{BHZ}, where the relation to discrete operators is discussed.
In fact, if we regard an infinitely extended solid whose ions are assumed 
to be fixed, then the positions are naturally distributed according 
to the points of a Delone set. Fixing an effective potential $v$ for
all the ions this leads us to consider the Hamiltonian
$$
H(\omega):=-\Delta + \sum_{x \in\omega}v(\cdot - x)\mbox{  in  }\D{R}^d,
$$
where $\omega\in \dell$. Let us assume, for simplicity that $v$ is bounded,
measurable and compactly supported.

In order to apply our analysis above, we need to introduce a suitable
topology on $\dell$. This can be done in several ways, cf. \cite{BHZ,
oamp}. We follow the strategy from the latter paper and refer to it 
for details (see the discussion in the appendix as well). The \emph{natural topology} defines a compact, completely
metrizable topology on the set of all closed subsets of $\D{R}^d$ for
which $\dell_{r,R}(\D{R}^d)$ is a compact, complete space. Moreover, the map
$$ H : \dell_{r,R}(\D{R}^d) \longrightarrow \F{S} (L^2 ((\D{R}^d)), \;\: \omega\mapsto H(\omega), $$
is continuous. This is a straightforward consequence of the following
lemma, which describes convergence w.r.t the natural topology.

\begin{lem} \label{conv}
  A sequence $(\omega_n)$ of Delone sets converges to $\omega\in\dell$
  in the natural topology if and only if there exists for any $l>0$ an
  $L>l$ such that the $\omega_n\cap U_L(0)$ converge to $\omega\cap
  U_L(0)$ with respect to the Hausdorff distance as $n\to\infty$.
\end{lem}

The \textit{Proof} of the lemma will be given in the appendix.  There,
we also discuss further features of the natural topology.
We say that a
Delone set $\rho$ is \textit{crystallographic} if $Per(\rho):=\{t\in \D{R}^d :
\rho = t+ \rho \}$ is a lattice.

\medskip

We are now in position to state the main application of this paper:
\begin{thm}\label{delone} Let $r,R>0$ with
$2 r\le R$ and $v$ be given such that there exist crystallographic $\gamma,
\tilde{\gamma} \in \dell_{r,R} $ with $\sigma(H(\gamma))\not=\sigma(H(
\tilde{\gamma}))$. Then
$$U:=(\sigma(H(\gamma))^\circ\setminus\sigma(H(
\tilde{\gamma})))\cup (\sigma(H(
\tilde{\gamma}))^\circ\setminus \sigma(H(\gamma)))$$ is nonempty and there
exists a  dense $G_\delta$-set  $\Omega_{sc}\subset \dell_{r,R}$ such that
for every $\omega\in \Omega_{sc}$ the spectrum of $H(\omega)$ contains
$U$ and is purely singular continuous in $U$.
\end{thm}

To prove the theorem, we need two results on extension of Delone
sets. To state the results we use the following notation.  For $S>0$
we define $Q(S) :=[-S,S]^d \subset \D{R}^d$. 

\begin{lem}\label{exteins}  
Let $r,R>0$ with $2 r\le R$ be given. Let $\omega \in \dell_{r,R}$ and
$S>0$ be arbitrary. Then, there exists a crystallographic $\rho \in
\dell_{r,R}$ with $\rho\cap Q(S) = \omega \cap Q(S)$.
\end{lem}
\begin{proof} 
Let $P : \D{R}^d \longrightarrow \D{R}^d / 2 (S+ R + r) \D{Z}^d =: \TT$
be the canonical projection. Note that the Euclidean norm  on $\D{R}^d$ induces
a canonical metric $e$ on $\TT$ with $e (P(x),P(y)) = \| x- y\| $
whenever $x,y\in \D{R}^d$ are close to each other.

Let $F_0 := P(\omega \cap Q(S+R))$. By assumption on $\omega$, we have
$e(p,q) \geq 2 r$ for all $p,q\in F_0$ with $p\neq q$. Moreover,
$\cup_{p\in F_0 } B_e (p,R) \supset P(Q(S))=:C$, where $B_e (p,R)$
denotes the ball around $p$ with radius $R$ in the metric $e$.

Adding successively points from $\TT \setminus C$ to $F_0$ we obtain a
finite set $F$ which is maximal among the sets satisfying
$$ e (p,q) \geq 2 r, \mbox{for all $p,q\in F$ with $p\neq q$.}$$
As any larger set violates this condition and $R\geq 2 r$, we infer
$$\cup_{p\in F} B_e (p,R) = \TT.$$
 Now, $\rho := P^{-1} (F)$ has the
desired properties. 
\end{proof}

\begin{lem}\label{extzwei}
Let $r,R>0$ with $2 r\le R$ be given. Let $\gamma,\omega \in
\dell_{r,R}$ and $S>0$ be arbitrary. Then, there exists a $\rho \in
\dell_{r,R}$ with
$$ \rho \cap Q(S) =\omega \cap Q(S) \;\:\mbox{and}\;\: \rho \cap (
\D{R}^d \setminus Q(S +  2R +r)) = \gamma \cap ( \D{R}^d \setminus Q(S
+ 2 R +r)).$$
\end{lem}
\begin{proof} Define 
$$ \rho' := ( \omega \cap Q(S + R) ) \cup ( \gamma \cap ( \D{R}^d
\setminus Q(S + R +r)) ).$$ Then, $\cup_{x\in \rho'} B(x,R) \supset
Q(S) \cup ( \D{R}^d \setminus Q (S+ 2R + r))$ and $\|  x -y\|  \geq 2 r
$ for all $x,y\in \rho'$ with $x\neq y$. Adding successively points
from $Q( S + 2 R +r) \setminus Q (S) $ to $\rho$ we arrive at the
desired set $\rho$. 
\end{proof}

\begin{proof}[Proof of Theorem \ref{delone}]
We let $U_1:=\sigma(H(\gamma))^\circ\setminus\sigma(H(
\tilde{\gamma}))$ and $U_2:=\sigma(H( \tilde{\gamma}))^\circ\setminus
\sigma(H(\gamma))$. Since $\gamma, \tilde{\gamma}$ are crystallographic, the
corresponding operators are periodic and their spectra are
consequently purely absolutely continuous and consist of a union of
closed intervals with only finitely many gaps in every compact subset
of the reals. Hence, under the assumption of the theorem  $U_1$
or $U_2$ is nonempty.  Thus, $U$ is nonempty.

\smallskip

We now consider the case that $U_1$ is nonempty. We will verify 
conditions (1)-(3) from Theorem \ref{thm2.1} above:

Ad (1): Fix $\omega\in\dell_{r,R}$. For $n\in \D{N}$ consider
$\nu_n:=\omega\cap Q(n)$. By Lemma \ref{exteins}, we can find a
crystallographic $\omega_n$ in $\dell_{r,R}$ with $\omega_n \cap Q(n)
= \nu_n$.  For given $L>0$ we get that $\omega_n\cap U_L(0)=
\omega\cap U_L(0)$ if $n$ is large enough. Therefore, by Lemma
\ref{conv}, we find that $\omega_n\to \omega$ with respect to the
natural topology. On the other hand,
$\sigma_{pp}(H(\omega_n))=\emptyset$ since the potential of
$H(\omega_n)$ is periodic. Consequently,
$$
\{\omega\in\dell_{r,R}| \sigma_{pp}(H(\omega))\cap U_1=\emptyset\}
$$
is dense in $\dell_{r,R}$.

Ad (2): To get the denseness of $\omega$ for which
$\sigma_{ac}(H(\omega))\cap U_1=\emptyset$, fix $\omega\in
\dell_{r,R}$. For $n\in \D{N}$ large enough, we apply Lemma
\ref{extzwei} to obtain $\omega_n\in\dell_{r,R}$ such that
$$
\omega_n\cap U_n(0)= \omega\cap U_n(0)\mbox{  and  }
\omega_n\cap U_{2n}(0)^c=\widetilde{\gamma}\cap U_{2n}(0)^c .
$$ In virtue of the last property, $H(\omega_n)$ and
$H(\widetilde{\gamma})$ only differ by a compactly supported, bounded
potential, so that
$\sigma_{ac}(H(\omega_n))=\sigma_{ac}(H(\widetilde{\gamma}))\subset
U_1^c$. In fact, standard arguments of scattering theory give that
$e^{-H(\omega_n)}-e^{-H(\widetilde{\gamma})}$ is a trace class operator. By
the invariance principle, the wave operators for $H(\omega_n)$ and
$H(\widetilde{\gamma})$ exist and are complete which in turn implies
equality of the absolutely continuous spectra. See, e.g., \cite{rs3},
Section XI.3 and \cite{jk}, Section 2 as well as \cite{scatt}, Corollary
of Theorem 2 for a much more general result.

 Again, $\omega_n\to \omega$ yields
condition (2) of Theorem \ref{thm2.1}.

Ad (3): This can be checked with a similar argument, this time with
$\tilde{\gamma}$ instead of $\gamma$.  More precisely, we proceed as
follows: Fix $\omega\in \dell_{r,R}$. For $n\in \D{N}$ large enough,
we apply Lemma \ref{extzwei} to obtain $\omega_n\in\dell_{r,R}$ such
that
$$
\omega_n\cap U_n(0)= \omega\cap U_n(0)\mbox{  and  }
\omega_n\cap U_{2n}(0)^c={\gamma}\cap U_{2n}(0)^c .
$$ In virtue of the last property, $H(\omega_n)$ and $H({\gamma})$
only differ by a compactly supported, bounded potential, so that
$\sigma_{ac}(H(\omega_n))=\sigma_{ac}(H({\gamma}))\supset U_1$. By
$\omega_n \to \omega$, we obtain (3) of Theorem \ref{thm2.1}.

Summarizing what we have shown so far, an appeal to Theorem
\ref{thm2.1} gives that 
$$
\{\omega\in\dell_{r,R}|\;  \sigma_{pp}(H(\omega))\cap U_1=\emptyset,
\sigma_{ac}(H(\omega))\cap U_1=\emptyset,  U_1\subset\sigma(H(\omega))\}
$$
is a dense $G_\delta$-set if $U_1$ is not empty. 

An analogous argument shows the same statement with $U_2$ instead of
$U_1$.  This proves the assertion if only one of the $U_i$, $i=1,2$,
is not empty. Otherwise, the assertion follows after intersecting the
two dense $G_\delta$'s.
\end{proof}
\begin{rem*}
The assumption of the above theorem combines non triviality
of $v$ and the existence of suitable crystallographic  $\gamma,\tilde{\gamma}$.

A simple way of ensuring this condition for $v\not=0$ of fixed
sign is to choose $R>2r$. By an argument as in the proof of Lemma
\ref{exteins} we can then find crystallographic $\gamma,\tilde{\gamma}$
such that $0\in\tilde{\gamma}\setminus\gamma$. The corresponding
periodic operators differ by a periodic potential with the same periodicity
lattice. In this case, the analysis of \cite{kss} can be applied,
showing that the spectra of $H(\tilde{\gamma})$ and $H(\gamma)$
differ.
\end{rem*}

\begin{appendix} 

\section{The natural topology and the proof of Lemma \ref{conv}} 
The purpose of this section is to comment on the natural topology and
to give a proof for Lemma \ref{conv}. To do so we use the description
of the natural topology via a one point compactification as given in
our article \cite{oamp}.

We need some notation.  Whenever $(X,e)$ is a metric space, the
Hausdorff distance $e_H$ on the compact subsets of $X$ is defined by
$$e_H (K_1,K_2):= \inf( \{ \epsilon >0: K_1\subset  
U_\epsilon(K_2)\:\;\mbox{and}\:\; K_2\subset U_\epsilon(K_1)\}\cup \{1\} ),  
$$ where $K_1,K_2$ are compact subsets of the metric space $(X,e)$ and
$U_\epsilon(K)$ denotes the open $\epsilon$-neighborhood around
$K$ w.r.t $e$. It is well known that the set of all compact subsets of $X$
becomes a complete compact metric space in this way whenever $(X,e)$ is
 complete and compact.

Using the stereographic projection
$$j : \D{R}^d\cup \{\infty\} \longrightarrow \D{S}^d:=\{x\in
\D{R}^{d+1} : \|  x\| =1\}$$ 
we can identify $\D{R}^d\cup \{\infty\} $
and $\D{S}^d$. We denote the Euclidean distance on $\D{S}^d$ by
$\rho$. 
 Denote by $\CalF:=\CalF (\RR^d)$ the set of closed
subsets of $\RR^d$. We can then define a metric $\delta$ on  $\CalF (\RR^d)$ by

$$\delta (F,G) :=\rho_H ( j(F \cup \{\infty\}) , j(G\cup\{\infty\}) ).$$ 

This makes sense since $j(F\cup \{\infty \})$ is compact in
$(\D{S}^d,\rho)$ whenever $F$ is closed in $\RR^d$. Moreover, it is not
hard to see that $\{ j(F\cup \{\infty\}) : F\in \dell_{r,R}\}$ is
closed within the compact subsets of $\D{S}^d$ and hence compact.

\medskip

After this general discussion, we can now give a 

\begin{proof}[Proof of
Lemma \ref{conv}]  

We start by proving the ``only if'' part: Let $\omega_n, \omega\in \dell$ be given with
$\omega_n\to \omega $ in $(\CalF, \delta)$. Fix $l>0$. Since $\omega$
is discrete, we find $L >l$ such that $d (\partial U_L (0),\omega)
=:\gamma >0$, where $d$ refers to the Euclidean distance.  There
clearly exists $C_L$ with

$$ \rho (j(x),j(y)) \leq 2 \|  x -y\| \leq C_L  \rho (j(x),j(y)) \:\;\:\;\:\hfill (*)$$ for all $x,y\in U_L (0)$.  Fix $\epsilon >0$ and use convergence
of $ \omega_n$ to $\omega$ in $\CalF$ to find $n_0\in \NN$ such that
$\delta (\omega_n,\omega)\leq \beta \epsilon$ for $n\geq n_0$
for $\beta$ small enough so that
$$ C_L \beta \epsilon \leq 
\min\{\epsilon,\gamma\}.$$ By definition of $\delta$ this means that
for every $x\in \omega \cap U_L (0)$ there is $x_n \in \omega_n$ such that 
$\rho(j(x),j(x_n))\leq  \beta \epsilon$. From $(*)$, we get that 
$$ \| x - x_n\|  \leq \frac{C_L}{2} \beta \epsilon \leq \frac{1}{2}
   \min\{\epsilon,\gamma\}$$ so that $x_n \in U_L (0)$ and $\omega
   \cap U_L (0) \subset U_\epsilon (\omega_n \cap U_L (0))$ for all
   $n\geq n_0$.

Conversely, given $x_n \in \omega_n \cap U_L (0)$ we find $x\in
\omega$ such that $\rho (j(x),j(x_n))\leq  \beta \epsilon$. Again, by
our choice of $\gamma$, $\beta$ and $j(x)\in U_L (0)$, this implies
$\omega_n \cap U_L (0) \subset U_\epsilon (\omega\cap U_L (0))$.

These considerations show that $ \omega_n \to \omega$ in
$(\CalF,\delta)$ implies that for all $l>0$ there exists $L>l$ with
$d_H (\omega_n \cap U_L (0), \omega \cap U_L (0))\longrightarrow 0$. This proves the ``only if'' part of the lemma.

\smallskip

We are now going to prove the ``if'' part of the lemma.  Assume that
$\omega_n$, $\omega$ are Delone sets satisfying the condition of the
lemma. We use a standard compactness argument to show that
$\omega_n\to \omega$ w.r.t. $\delta$. Choose an arbitrary subsequence
$( \omega_{n_k})$ of $(\omega_n)$.  By compactness, there is a
subsequence $(\omega_{n_{k_l}})$ converging to some
$\widetilde{\omega}$ w.r.t. $\delta$: By the first part of the proof,
for every $l>0$, we then find $L>0$ with $d_H (\omega_{n_{k_l}} \cap U_L
(0), \widetilde{\omega}\cap U_L (0))\longrightarrow 0$. This implies
that $\omega = \widetilde{\omega}$. Therefore, every subsequence of
$(\omega_n)$ has a subsequence converging to $\omega$ w.r.t. $\delta$.
Thus, the sequence $(\omega_n)$ itself converges to $\omega$
w.r.t. $\delta$. This finishes the proof of the lemma.
\end{proof}

\medskip

The proof of the lemma shows effectively that $\omega_n\to \omega$ in
$(\CalF,\delta)$ if and only if the following two conditions hold:
\begin{itemize}
\item[(i)] For every $x\in \omega$, there exists $(x_n)$ with $x_n \in
\omega_n$ for every $n\in \NN$ and $x_n \to x$.
\item[(ii)] Whenever $(x_n)$ is a sequence with $x_n \in \omega_n$ for
every $n\in \NN$ and $x_n \to x$, then $x\in \omega$.
\end{itemize}

\end{appendix}

%-------------------------------------------------------------%

%-------------------------------------------------------------%
\end{document}